# Electroisomerization Blinking of an Azobenzene Derivative Molecule

Sylvie Godey, Hugo Therssen, David Guérin, Thierry Mélin, Stéphane Lenfant*

*Univ. Lille, CNRS, Univ. Polytechnique Hauts-de-France, UMR 8520 - IEMN - Institut d'Electronique de Microélectronique et de Nanotechnologie, F-59000 Lille, France.*
*Corresponding Author : stephane.lenfant@univ-lille.fr*

**Abstract**

We report here the reversibility and bistability of the switching behavior in an azobenzene derivative induced by the bias applied by a Scanning-Tunneling Microscopy (STM) tip, at low temperature and in ultra-high vacuum environment. This *cis*-to-*trans* and *trans*-to-*cis* switching were observed during STM imaging in either polarity at +2V or -2V, on a sub-second time scale. This results in a blinking effect visible on STM images, corresponding to the reversible switching of the azobenzene molecule under the applied STM bias through an electric field induced process.

**1. Introduction**

In recent years, there has been a growing interest in photo-switchable molecules[1], driven by their potential applications in various fields such as biology[2], nanomedicine[2], solar energy storage[3], information storage[4,5], nanomachines[6,7,8], smart surfaces[9], photo-controllable devices[10] or molecular electronics[11,12,13,14]. These photochromic molecules exhibit a common property: they can reversibly switch between two (or more) stable states when exposed to light with a specific wavelength. The transition induced by external light stimuli causes a modification in the molecular structure and its associated electronic or magnetic properties. Among the wide variety of photochromic molecular switches, azobenzene and its derivatives have received important attention due to their straightforward molecular structure and their robust and repetitive switching capabilities[15]. The azobenzene molecule and its derivatives are well known to switch reversibly between two isomeric states by photoisomerization: the nearly planar *trans*-isomer and the three-dimensional *cis*-isomer. The *cis*-to-*trans* isomerization was induced by exposure to visible light with a wavelength over 400 nm, and the *trans*-to-*cis* isomerization by exposure to UV illumination with wavelengths between 300 to 400 nm[6,15]. Scanning-probe microscopies such as scanning tunneling microscopy (STM) have been used as ideal tools to observe and manipulate such molecular switches on metallic surfaces[16]. The azobenzene isomerization has been extensively investigated in liquid and in gas phase[15], but also by STM for various azobenzene derivatives deposited onto different substrates such as Au(111)[17,18,19,20,21,22,23,24,25], Au(100)[24], Cu (111)[24,26], NaCl/Ag(111)[27] or GaAs (110)[28]. The Au(111) surface remains the most commonly employed surface for observing isomerization. Indeed, by comparing the three surfaces Au(111), Cu(111) and Au(100), electroisomerization was solely observed on the Au(111) surface, which is attributed to a weaker molecule-substrate interaction compared to the other surfaces, as indicated by the authors[24]. This makes the Au(111) a model surface for these studies. It has also been established that the isomerization of the azobenzene molecule can be induced by STM[27,17,19,23, 24,26], by applying voltage pulses with the STM tip at a controlled tip-surface distance. The STM also gives us perfect control over where the pulse is applied, allowing us to select the individual molecule to be isomerized.[17,19,27]

Here we report, using ultra-high vacuum (UHV) STM investigations mainly at 4 K on azobenzene derivative islands deposited on Au(111), that electroisomerization occurs reversibly during STM imaging at low bias, without introducing voltage pulses as reported in the previous studies mentioned above, and at a short (i.e. sub-second) timescale. The reversible electroisomerization process is simply and directly observed during STM imaging in both polarities, leading to the observation of the blinking effect associated to the rapid modification of the isomer under the electric field.

In order to promote the isomerization of the azobenzene part, we designed an azobenzene derivative specifically to reduce its interaction with the surface[20]. The 3,3',5,5'-tetra-tert-butylazobenzene molecule in Figure 1 (hereafter named TBA) has four lateral *tert*-butyl-groups which act as "spacer legs" to reduce the electronic coupling between the active part of the molecule (azobenzene) and the metal surface[20]. This TBA molecule is known to switch reversibly between its two isomeric *trans*-TBA and *cis*-TBA states by photoisomerization either under UV irradiation (from *trans*-TBA to *cis*-TBA), under blue light irradiation (from *cis*-TBA to *trans*-TBA)[20,25,22,21,28] (Figure 1), or by electroisomerization under an electric field induced by a STM tip[23,24]. The latter occurs with an applied electric field between 0.1 to 0.7 V/Å between the STM tip and the gold surface (typically 30 s voltage pulses with a bias comprised between -5 to 8 V at a controlled tip-sample distance comprised between 0.4 to 1.6 nm)[23]. For these authors, the electroisomerization was also observed with higher tip-surface distance of 3.6 nm with a voltage pulse of 6.8 V where no tunneling current is flowing through the sample-tip gap[23]. From these observations, the authors concluded that the *trans*-to-*cis* and *cis*-to-*trans* electroisomerizations are mainly caused by the electric field and not by tunneling current[23]. We can notice, as indicated in Figure 1, that the *cis* isomer can thermally relax to the *trans* isomer. This is because the *trans* isomer is the more stable form of the molecule[29].

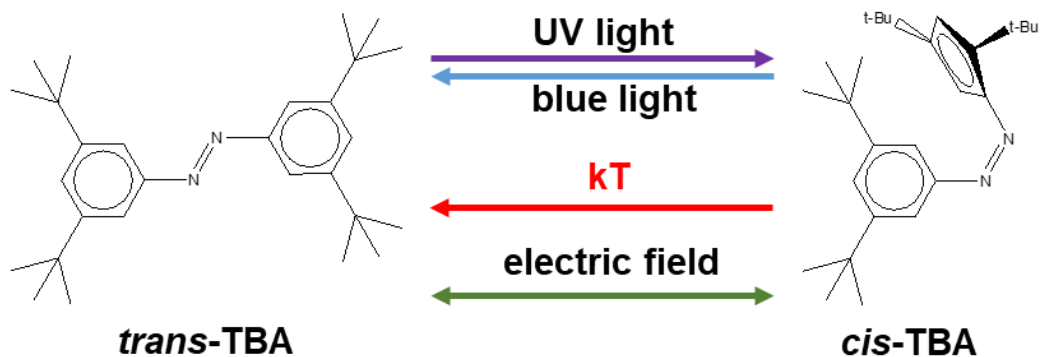

**Figure 1.** *Chemical structure of the-3,3',5,5'-tetra-tertbutyl-azobenzene molecule (TBA) in the two reversible isomers:* trans-*TBA and* cis-*TBA. The isomerization between the two forms,* trans-*TBA and* cis-*TBA, of the TBA molecule occurs in particular through exposure to*

*light (photoisomerization) or under the influence of an electric field (electroisomerization). The* cis *molecule can also thermally relax to the stable* trans *form.*

**2. Method**

TBA molecule synthesis has been here performed following the procedure described by Alemani et al.[23], and confirmed via Nuclear Magnetic Resonance (NMR) and UV-Visible spectroscopy (refer to $^1$H and $^{13}$C-NMR spectra in CDCl$_3$ and UV-visible spectrum in hexane solution in the section 2 of the Supplementary Information). A monocrystalline Au(111) surface bought from Mateck (Germany) was prepared by multiple cycles of Ar sputtering and thermal annealing at about 700 K. TBA molecules were deposited at room temperature onto the freshly prepared gold surface by thermal evaporation at 368 K for about 50 s. STM experiments were conducted at 4 K using a Joule-Thomson scanning-probe microscope (SPECS JT AFM/STM, Berlin) with a base pressure of $10^{-10}$ mbar. STM characterizations were realized using tungsten tips mounted on Kolibri sensors (SPECS, Berlin). The Kolibri sensor (with a 100 pm peak-to-peak oscillation amplitude) was used here only in the STM mode. During the measurements, the tip of the Kolibri was never in contact with the surface. The samples have been used for preliminary tests upon optical illumination[21], before conducting the electrical switching experiments described in this Article. The resulting topographic images presented in this work were processed and displayed using WSxM 5.0 free software[30], with the following functions: Global plane, Flatten, Zoom and Z control (see examples in section 7 of the Supplementary Information).

**3. Results and discussion**

After the deposition of TBA molecules, due to their mobility on the clean Au(111) surface, STM images (Figure 2a) reveal the formation of (i) TBA islands on gold terraces; (ii) individual TBA molecules located at the elbows of the Au(111) herringbone reconstruction (circles in Figure 2a); and (iii) TBA molecules along the step edges. Islands generally consist of parallel rows of molecules with a short-range order, where neighboring molecules are aligned with each other (Figure 2b). TBA molecules are imaged by STM as four distinct lobes attributed to the four *tert*-butyl groups or "spacer legs". The azobenzene part of the TBA molecule is not visible in STM[21,23]. The apparent height for the lobes is measured as 0.25 ± 0.02 nm. The arrangement of the four lobes and their height observed are systematically associated with the *trans*-TBA isomer on the metal surface, indicating a planar geometry of the TBA molecule.[20,12,24,14,22,21] The *trans*-TBA isomer is the most energetically stable form[29,31], explaining why the isomer is observed after the TBA evaporation on gold surfaces. In these *trans*-TBA island lattices, a unit cell is proposed in Figure 2b. The unit cell has side length of 1.06 ± 0.04 nm and 1.42 ± 0.05 nm with an opening angle of 107 ± 1°. These values are in full accordance with those measured on similar molecules (TBA with a methoxy group) as imaged by AFM on calcite with the reported values of 1.04 ± 0.04 nm, 1.56 ± 0.03 nm and 106.9 ± 0.5° respectively.[32] These authors also associated this unit cell with the TBA molecule in the *trans* isomer.

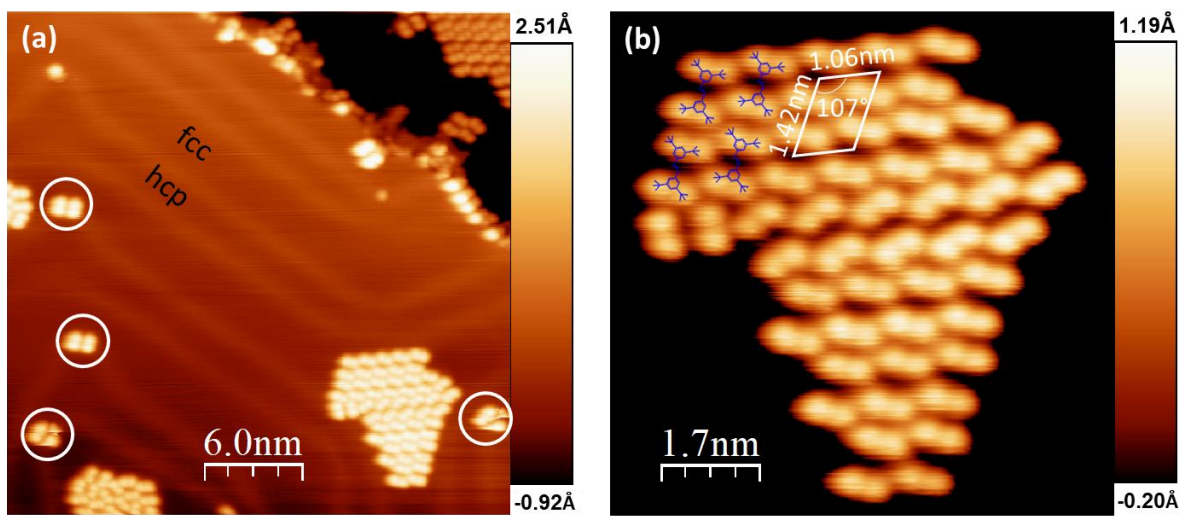

**Figure 2.** *(a) STM constant-current image of TBA after deposition on Au(111), 30 nm x 30 nm, I = 10 pA, V = -1.5 V at 4 K with the Z-scale on the right. Circles correspond to single TBA molecules localized at the elbows of the Au(111) reconstruction. The characteristic Au(111) (22 x √3) "herringbone" fcc and hcp[31] patterns are indicated. (b) Zoom on a molecule island composed of organized* trans-TBA *molecules, 8.7 x 8.7 nm², I = 10 pA, V = + 1.5 V with the Z-scale on the right. The unit cell (white parallelogram) of the* trans-TBA

*molecule local organization is presented. TBA molecules are identified in blue only as a guide for the eyes.*

Successive STM images acquired on a *trans*-TBA island freshly deposited, consisting of approximately 180 molecules, are presented in Figures 3a, 3b, 3c and 3d using a sample voltage of -1 V, -2 V, -2 V and -1 V respectively. For the first image at -1 V (Figure 3a), the observed island is comparable to the one presented above (Figure 2) in terms of the molecule shape and organization. The image at -2 V reveals a modification in the morphology compared with the image at -1 V, particularly with the appearance of approximately 31 bright events within the island of ~ 180 molecules (corresponding of a switching ratio of 17%). The number of these bright events increases by 16 in the following images acquired at -2 V (corresponding of a switching ratio of 11% - Figure 3c) and by 9 in the image acquired at -1 V (corresponding of a switching ratio of 7% - Figure 3d). This modification in the isomerization of some molecules in the final image at -1V (Figure 3d) is explained by the electroisomerization of the molecule after being imaged by STM at -2V in Figure 3c. As mentioned by Alemani et al.[23], this electroisomerization can occur over lateral distances of up to 8 x 8 nm² from the position of the STM tip. The image remained unchanged in subsequent images acquired at V = -1V (images not shown in the main Article, see Supplementary Information in Figure SI-5). Generally, the shape of the bright events is not perfectly round in the images acquired at -2 V; in some cases, only a few bright lines constitute the bright event, which indicates an instability of the bright state. In contrast, in the image acquired at V = -1 V (Figure 3d), the shape of the bright events is perfectly round and remains so in the subsequent images acquired at -1 V (see Supplementary Information in Figure SI-5).

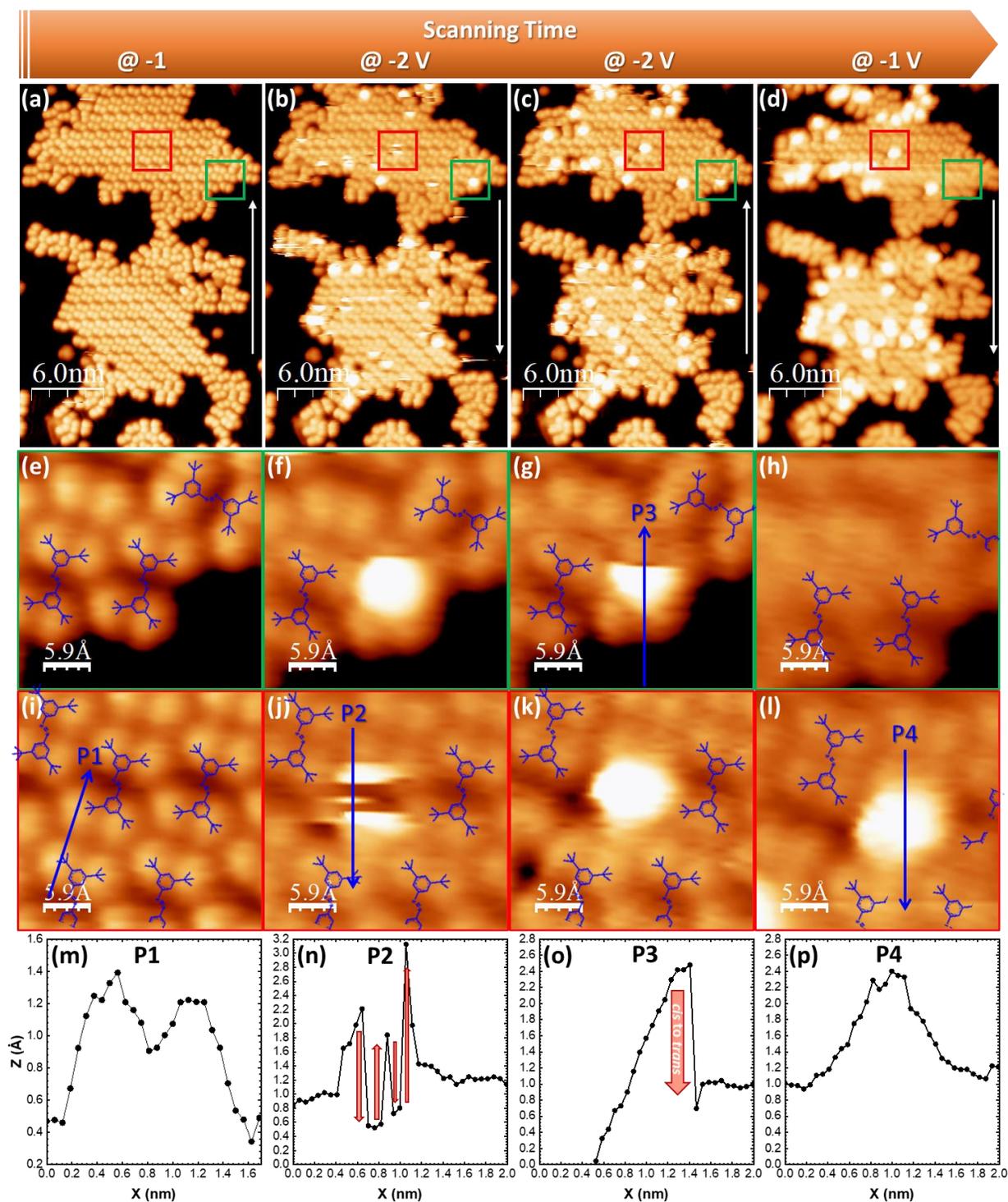

**Figure 3.** *Four successive 20 x 30 nm² images of a TBA island were acquired by STM (I = 5 pA) at V = -1 V in Y-scan upward (a); V = -2V in Y-scan downward (b); V = -2 V in Y-scan*

*upward (c); and V = –1 V in Y-scan downward (d). White arrows indicate the Y scanning direction. Each image acquisition time was 6 min 30 s. The two sequences (e), (f), (g), (h) and (i), (j), (k), (l) are zoomed images (size 2.9 x 2.9 nm²) of the previous images (a), (b), (c), (d), corresponding to the green and red squares respectively. Four topographic profiles (P1, P2, P3 P4) represented by blue arrows in images (i), (j), (g), (l) are plotted in (m), (n), (o), (p) respectively. The direction of the gray arrows indicates the scanning direction along the Y-axis. TBA molecules are identified in blue only to guide the eye.*

Two zoomed STM images on a few molecules (represented by red and green squares in Figure 3a) are shown for successive voltages in Figures 3e to 3l. The two zoomed images acquired at -1 V (Figures 3e and 3i) are similar to those in Figure 2, showing the presence of the four lobes corresponding to the molecule in the *trans* configuration. During imaging at -2 V, bright events are observed in both sets of zoomed images. However, at the end of the voltage sequence, the zoom shown by the green square leaves the molecule unchanged (Figure 3h), while the zoom shown by the red square results in a round spot instead of the molecule with the four legs. The formation of these bright spots only appears during STM imaging with a polarization of -2 V and is never observed during STM imaging at -1 V (Figure SI-5). They are associated with an increase in apparent height of 0.23 ± 0.02 nm (profile in Figure 3p), which is characteristic of the formation of a *cis*-TBA isomer on the gold surface[20,12,24,14,22,21]. Therefore, in the first case (green square), the molecule remains unchanged in the *trans* conformation, whereas in the second case (red square), the molecule switches to the *cis* conformation at the end of the procedure. This *trans*-to-*cis* isomerization observed here was clearly caused by the application of a more negative applied voltage -2 V during STM imaging. It is noteworthy that the same switching behavior was also observed during imaging at a substrate voltage of +2 V (see the movie of STM imaging during 15 hours at 77K in Supplementary Information), indicating that the isomerization under STM imaging with bias higher than |2|V occurs at both polarities.

In the intermediate STM images obtained at V = -2V, instabilities with abrupt changes in apparent height were observed as evidenced for example by the profiles P3 and P2 (see Figure 3o and 3n respectively) obtained from Figures 3g and 3j. In the P3 profile (Figure 3o), during the scan, we first observe a bright zone corresponding to the *cis*-TBA molecule; then, in one line, the height decreases by 0.18 ± 0.02 nm, resulting in the final situation corresponding to the *trans*-TBA molecule. This abrupt modification between two consecutive scan lines, at a sub-second time scale (each line is acquired in 0.36 s), is associated with *cis*-to-*trans* electroisomerization. This value of 0.36 s for the electroisomerization time observed here can be compared to the time reported by Alemani et al.[23] These authors applied up to 9 pulses of 20–30 s to observe the *trans*-to-*cis* isomerization, corresponding to a total duration, up to 270 s, much longer than the times observed here.

Based on this observation, the electroisomerization is estimated to occur on a sub-second time scale. These abrupt changes between two consecutive lines, all associated with an isomerization change, are frequently observed during imaging at -2V, as seen in Figures 3f, 3g, and 3j. For Figure 3j, four such changes are observed (other examples are also presented in SI section 6). In the P2 profile (Figure 3n), four abrupt modifications were observed: a decrease of 0.15 ± 0.02 nm and an increase of 0.19 ± 0.02 nm during imaging, associated with *cis*-to-*trans* and *trans*-to-*cis* electroswitching, respectively. Between these two higher modifications, a weak peak with an amplitude of 0.11 ± 0.02 nm was observed, associated with *trans*-to-*cis* electroisomerization followed by *cis*-to-*trans* electroisomerization in the next scanned line. In this Figure 3n, electroisomerization is also observed along a single line during the STM scanning, allowing us to estimate an upper time for the electroisomerization process of approximately 6 ms (see details in section 5 of the supplementary information). These rapid modifications during imaging in the P2 profile lead to blinking of the molecule. This blinking is associated with switching between the two isomers, demonstrating the reversibility and bistability of the isomerization for the same bias. In these successive STM images, we directly observe the switching of the TBA molecule without application of repetitive voltage pulses at a fixed tip-sample distance, as realized elsewhere[23,24].

During the acquisition of this STM images, the risk of molecule transport by the STM tip is always possible. The molecule can attach to the STM tip and be deposited elsewhere on the surface. This would result in the molecule appearing or disappearing in successive images acquired at the same location. However, in all the images presented in the article, as well as in all the images obtained in this study, we have not observed any changes in the number of molecules during successive image acquisitions at the same location: no clear disappearance or appearance of molecules. This last point allows us to exclude any possible transport of molecules by the STM tip during imaging and, in particular, excludes the possibility that the bright spots are due to the deposition of a molecule on the surface.

The electroisomerization of the azobenzene derivatives can be explained by two distinct processes: (i) an electronically induced process at a low sample-tip distance with a tunneling current flow; or (ii) an electric field induced process at a higher sample-tip distance without tunneling current flow[23, 24]. In the first case, for the electronically induced process, the STM tip was positioned at the center of the azobenzene group (the double N=N bond) at a fixed tip-sample distance. Electrons were then injected into the azo group of the *trans* isomer by increasing the voltage bias. At a well-defined voltage bias, a modification of the STM image[19,27] or a drop in the tunneling current was observed[17]. These events are associated with a *trans*-to-*cis* switching. In the case of the current drop, this drop was only observed at negative bias[17], and switching was not observed in both polarities. In the second case, for the electric field induced process, authors applied 30 s voltage pulses on the TBA island, at a tip-surface distance ranging from about 0.4 to 1.6 nm, with a voltage bias ranging from -5 to 8 V.[23] Reversible isomerization was observed and associated with an electric field induced process because this isomerization was also observed for large tip-surface distances where no tunneling current is flowing[23]. According to these authors, the electroisomerization was caused by the electric field located between the STM tip and the sample surface at a fixed voltage.

In our case, the electroisomerization effect was only observed during imaging at -2 V but not during imaging at -1V. For both polarizations, the current setpoint was fixed at the same value (5pA), suggesting that the electroisomerization observed here is primarily an electric field-induced process caused by the increase in the value of the electric field while shifting the bias from -1 V to -2 V.

We can estimate the electric field necessary to induce the *trans*-to-*cis* isomerization by assuming that, for a current setpoint fixed between 5 - 10 pA, the tip-surface distance is around 1 nm. The switching observed at a STM bias of +2V, corresponding here to an electric field of approximately $2 \times 10^9$ V/m. This value is consistent with the range of $1 \times 10^9$ to $7 \times 10^9$ V/m reported by Alemani et al.[23], which was determined to induce the *trans*-to-*cis* isomerization. The contribution of the *cis*-isomer dipole to the molecular switch activation energy can be estimated in a simplified approach, as from the interaction of the electric field ($E$) with the molecular dipole ($\mu$). Using the dipole moment value in vacuum of 3.6 Debye for the *cis*-isomer reported in the literature[33,34], the potential gain in electrostatic energy (approximated by $E.\mu$) can be estimated at ~ 0.15 eV, which should be further screened due to the substrate metallic behavior. It will therefore be much lower than the activation energy of about 1 eV reported by Wolf and al.[35] for the TBA molecule on Au(111) surface. This is consistent with the fact that the switching behavior can be observed independently of the positive or negative tip polarity[23]. Its suggests that the switching barrier is rather governed by the molecule/surface interaction, and potentially polarizability effects under the tip electric field as previously discussed by Füchsel et al.[34].

From an energetic point of view, Figure 4 schematically illustrates the scheme of the potential energy for the reversible switching process. The two isomer states of the TBA molecule are depicted as two potential wells, with the *trans* state having a lower energy[29,31,36,35]. In the liquid or gas phase, the *trans*-state is 0.6 eV more stable than the *cis* state, with an activation energy of around 1 eV between these two states.[35] It is well established that the reversible photoisomerization with UV and blue irradiations and the *cis*-to-*trans* thermal activated isomerization can be explained by this two potential wells approach.[37,31] To photoswitch the molecule, it is necessary to overcome this activation energy. This is achieved by bringing the molecule to an excited state through illumination with photon energy ranging from 2.6 eV (visible) to 3.4 eV (UV). Afterward, the molecule relaxes and is converted into the other form. Based on the studies on the electroisomerization of azobenzene derivatives, the two potential wells scheme can be completed by an additional mechanism based on electric field induced isomerization. This electroisomerization is reversible (*cis*-to-*trans* and *trans*-to-*cis*), occurs at STM bias around V >= +2 V and V <= -2 V for a current set-point fixed at 5 pA, and operates down to the millisecond timescale. In a first approach, we can estimate that a STM bias of |V| = 2 V contributes an additional energy of 2 eV for the molecule. While this energy value may not be sufficient to excite the molecule, as achieved with illumination, it is sufficient to overcome the activation energy barrier of 1 eV and switch to the other isomer.

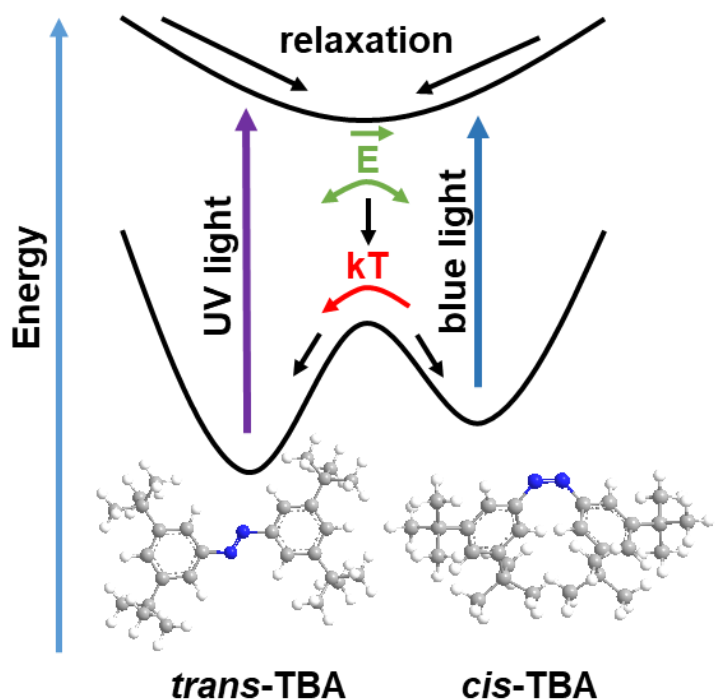

**Figure 4.** *Scheme of reversible photoisomerization induced by UV or blue light, reversible electroisomerization induced by the electric field, and the non-reversible* cis-*to*-trans *isomerization induced by heat (kT).*

## 4. Conclusion

In conclusion, we have successfully achieved organized TBA molecule islands deposited on Au(111) terraces and characterized them using STM in UVH at 4 K. During STM imaging at a bias voltage of -2 V, we observed rapid blinking with a duration on the sub-second time scale, corresponding to reversible isomerization of the TBA molecule under the influence of a higher electric field applied by the STM. The control of isomerization in molecular switches using an electric field, instead of the usual light-based methods, paves the way for the integration of these photochromic molecules into electronic devices, promising exciting prospects for future applications.

**Supplementary material**

See the supplementary material for the synthesis, $^1$H and $^{13}$C-NMR spectra of the TBA molecule in CDCl$_3$, UV-visible spectrum in hexane solution, STM images at -1 V after the sequence in Figure 3, the observation of the TBA electroisomerization during STM imaging along a line, other examples of electroisomerization observed during STM imaging à -2 V, examples of STM images processing with WSxM software, and the movie of STM imaging during 15 hours at +2 V and at 77K.

**Data availability**

The data that support the findings of this study are available from the corresponding author upon reasonable request and available on request via the website https://recherche.data.gouv.fr/en.

**Acknowledgements**

The IEMN scanning-probe microscopy characterization facilities (PCMP-PCP) are partly supported by the French Network Renatech. This work has been financially supported by the French National Research Agency (ANR), project SPINFUN ANR-17-CE24-0004.